\documentclass[prd,amsmath,twocolumn,nobibnotes,nofootinbib,showpacs,superscriptaddress]{revtex4}

\usepackage{bm}
\usepackage{graphicx}
\usepackage[hyperfootnotes=false,colorlinks,bookmarks]{hyperref}

\DeclareMathOperator{\Var}{Var}

\begin{document}

\title{The cosmic variance of $\Omega$}
\author{T. P. Waterhouse}
\email{tpw@cantab.net}
\affiliation{Perimeter Institute for Theoretical Physics, 31 Caroline Street North, Waterloo, ON, N2L 2Y5, Canada}
\affiliation{Department of Physics \& Astronomy, University of British Columbia, Vancouver, BC, V6T 1Z1, Canada}
\author{J. P. Zibin}
\email{zibin@phas.ubc.ca}
\affiliation{Department of Physics \& Astronomy, University of British Columbia, Vancouver, BC, V6T 1Z1, Canada}
\date{10 April 2008}

\begin{abstract}
How much can we know about our Universe?  All of our observations are restricted to a finite volume, and therefore our estimates of presumably global cosmological parameters are necessarily based on incomplete information.  Even assuming that the Standard Model of cosmology is correct, this means that some cosmological questions may be unanswerable.  For example, is the curvature parameter $\Omega_K$ positive, negative, or identically zero?  If its magnitude is sufficiently small, then due to cosmic variance no causal observation can ever answer that question.  In this article, we first describe the gauge problems associated with defining the cosmic variance of cosmological parameters, then describe a solution involving the use of parameters defined on the surface of last scattering, and finally calculate the statistical variance of ideal measurements of the matter, radiation, and curvature density parameters.  We find that $\Omega_K$ cannot be measured to better than about $1.5\times10^{-5}\ (1 \sigma)$, and that this limit has already begun to decrease due to the flattening effect of dark energy.  Proposed $21\,\text{cm}$ hydrogen experiments, for example, make this limit more than just a theoretical curiosity.
\end{abstract}


\maketitle

\section{Introduction}
The Standard Model of Cosmology is a remarkable scientific achievement.  The combination of general relativity with statistical homogeneity and isotropy yields a model which currently requires only about a half dozen parameters~\cite{Scott} in order to provide a successful description of the real Universe.

Assuming this Standard Model is correct, the physical Universe is larger (possibly infinitely so) than the observable Universe contained within our particle horizon.  Critically, the numbers that we consider to be ``cosmic parameters'', for example the total matter density parameter $\Omega_\text{m}$, are subject to sample variance, since we estimate them by making observations within our own causal patch.  This difference between a specific realization and the ensemble average is called \emph{cosmic variance}.  It is most familiar in the context of the observed values of the multipoles of the microwave sky.  Indeed, the cosmic variance in those multipoles places corresponding limits on our ability to constrain the cosmological parameters using the observed multipoles.

This decade's ``era of precision cosmology'' has seen these parameters measured to impressive precision~\cite{WMAP5}.  Today the uncertainty in our estimates of cosmological parameters is dominated by the limitations of our instruments, but cosmic variance will become dominant in the distant future if the sensitivity of our instruments continues to improve.  In particular, experiments to map the distribution of matter at high redshift using $21\,\text{cm}$ hydrogen line observations may be able to constrain the curvature parameter $\Omega_K$ to within $\sim10^{-4}$~\cite{0802.1710}.

In this article, we discuss cosmic variance as it relates to the density and curvature parameters $\Omega_\text{m}$, $\Omega_\text{r}$, and $\Omega_K$.  Ref.~\cite{Zibin} addresses some related issues regarding the cosmic microwave background dipole and monopole.

\subsection{Overview}
The total matter content of a homogeneous and isotropic universe is characterized by the dimensionless time-dependent parameter
\begin{equation}
\label{omegamdef}
\Omega_\text{m}\equiv\frac{8\pi G}{3H^2}\rho_\text{m},
\end{equation}
where $H$ is the Hubble rate and $\rho_\text{m}$ is the energy density of matter (dark matter and baryons).  The radiation density parameter $\Omega_\text{r}$ is defined analogously, in terms of $H$ and radiation density $\rho_\text{r}$ (photons and neutrinos).

Likewise, spatial curvature is characterized by
\begin{equation}
\Omega_K \equiv -\frac{{}^{(3)}\!R}{6H^2},
\label{omegakdef}
\end{equation}
where ${}^{(3)}\!R$ is the Ricci curvature scalar of the homogeneous spatial slices.  Thus $\Omega_K$ is determined by the ratio of the Hubble length to the spatial curvature length.  The parameter $\Omega_K$ is also time-dependent:  Since the spatial curvature decays like $a^{-2}$ in an expanding universe, for scale factor $a$, we have
\begin{equation}
\Omega_K(t)a^2(t)H^2(t) = \text{const}.
\label{omegaktdep}
\end{equation}
This time dependence encodes the familiar behavior that $\Omega_K = 0$ is unstable in an expanding matter- or radiation-dominated universe, leading to the flatness problem of hot big bang cosmologies, but \emph{stable} in an inflating universe, leading to the widely accepted resolution of that problem.

Indeed, inflation is usually thought to last much longer than is needed to flatten the Universe to within observed bounds on $\Omega_K$.  (Current constraints are $-0.0175 < \Omega_K < 0.0085$ at $95\%$ confidence~\cite[Table 2]{WMAP5}.)  Therefore it is expected that the Universe is essentially flat on scales much larger than the current radius of the last scattering surface, $r_\text{LS}$.  Equivalently, the ensemble average (over realizations of the primordial fluctuations) of the spatial curvature averaged over our observable volume is expected to be extremely small.

However, as it drives the spatial curvature towards zero, inflation also generates the near-scale-invariant spectrum of primordial fluctuations required to form structure~\cite{Liddle_Lyth}.  The primordial comoving curvature perturbations carry a characteristic dimensionless amplitude of order $10^{-5}$, which suggests that even an ideal measurement of spatial curvature will be subject to irreducible random noise, or cosmic variance, of order $\Omega_K \sim 10^{-5}$ (see, e.g., Knox~\cite{knox06}).  For adiabatic fluctuations, the Einstein energy constraint equation then suggests that the parameters $\Omega_\text{m}$ and $\Omega_\text{r}$ should be subject to cosmic variance of similar relative size.  Given a primordial spectrum, it is possible to use linear perturbation theory to quantify these expectations and calculate the cosmic variance of, and hence the best possible constraints on, the density parameters $\Omega_\text{m}$, $\Omega_\text{r}$, and $\Omega_K$.

To do this, we must generalize Eqs.~(\ref{omegamdef}) and~(\ref{omegakdef}) to the case of universes that depart from exact homogeneity and isotropy.  The most natural approach is to take these equations to define \emph{local} parameters $\Omega_\text{m}(\bm{x},t)$, $\Omega_\text{r}(\bm{x},t)$, and $\Omega_K(\bm{x},t)$ in terms of similarly local densities $\rho_\text{m}(\bm{x},t)$ and $\rho_\text{r}(\bm{x},t)$, curvature ${}^{(3)}\!R(\bm{x},t)$, and Hubble rate $H(\bm{x},t)$,\footnote{We define $H(\bm{x},t)$ to be the expansion rate of the worldlines comoving with matter (cold dark matter plus baryons), i.e.\ the worldlines which observe zero total matter momentum density.} and to take $\Omega_\text{m}$, $\Omega_\text{r}$ and $\Omega_K$ to be the spatial averages of $\Omega_\text{m}(\bm{x},t)$, $\Omega_\text{r}(\bm{x},t)$, and $\Omega_K(\bm{x},t)$, respectively, for a fixed time $t$.  However, since we no longer have homogeneous hypersurfaces naturally singled out, we are free to vary the choice of time coordinate.  For example, we could choose the slices of constant $t$ to be spatially flat, in which case ${}^{(3)}\!R$ and hence $\Omega_K$ trivially vanish everywhere!  This inevitable gauge ambiguity can be defeated by fixing the slicing by a \emph{physical} criterion.  Since we will wish to examine the largest observable slice in order to obtain the best constraints, we will choose to work on the hypersurface of last scattering, which is defined locally by the condition that the radiation energy density be some constant value.

\subsection{Structure, assumptions, and conventions}
The structure of this paper is as follows:  In section~\ref{CPT}, we introduce the basics of cosmological perturbation theory, describe the problem of gauge ambiguity in defining cosmological parameters, and propose a solution involving choosing new parameters defined on the hypersurface of last scattering.  In section~\ref{PLS}, we derive the rule for transforming from one hypersurface to another and the relationship between density perturbations and density parameter perturbations, for matter, radiation, and curvature.  In section~\ref{numerical_section}, we numerically compute the cosmic variance of ideal measurements of the matter, radiation, and curvature density parameters we have defined.  In section~\ref{directOmegaKsec}, we employ an alternate method to compute the variance of the curvature parameter, highlighting the importance of the definition of curvature.  In section~\ref{IaD}, we adapt the results of the previous sections to give the variance of $\Omega_K$ in the present era, allowing comparison to observations and showing that it is impossible to measure $\Omega_K$ to better precision than $1.5\times10^{-5}$.

Motivated by inflation and by current data, we assume vanishing \emph{background} spatial curvature and strictly adiabatic perturbation modes.  Throughout this paper we use the convention $c=1$, and all distances are comoving.  Of course, we also assume the Standard Model of cosmology to be correct.  This is sensible, since we are trying to find minimum theoretical uncertainties, and if the real map of the Universe has dragons beyond our particle horizon, then our lower bound will certainly still hold.

\section{\label{CPT}Cosmological perturbation theory}
The Friedmann-Lema\^itre-Robertson-Walker model describes a universe which is homogeneous and isotropic.  Cosmological perturbation theory is the means to generalize this model to the relaxed requirement of statistical, rather than exact, homogeneity and isotropy.  We begin with a brief summary of gauge choice in perturbation theory.  See, e.g., Mukhanov~\cite[section 7.1]{Mukhanov} for more details.

We can introduce perturbations to an FLRW universe by writing the metric as
\begin{displaymath}
g_{\mu\nu}=g_{\mu\nu}^\text{FLRW}+\delta g_{\mu\nu}.
\end{displaymath}
Here $\delta g_{\mu\nu}$ is a symmetric 4-by-4 matrix, not a tensor, and it is convenient to decompose it as
\begin{equation}
\label{delta_g}
\delta g_{\mu\nu}=a^2\left(
\begin{array}{cc}
2\phi&B_{,i}+S_i\\*
B_{,j}+S_j&2\psi\gamma_{ij}+2E_{,ij}+F_{(i,j)}+h_{ij}
\end{array}
\right),
\end{equation}
breaking down the 10 degrees of freedom of $\delta g_{\mu\nu}$ into
\begin{itemize}
\item four scalar degrees in the form of four scalar fields $\phi$, $\psi$, $B$, and $E$, which give rise to structure in the Universe;
\item four vector degrees in the form of two divergence-free spatial vectors $F_i$ and $S_i$, which decay quickly as space expands;
\item and two tensor degrees in the divergence-free, trace-free spatial tensor $h_{ij}$, which give rise to the two polarizations of gravitational radiation.
\end{itemize}

The transformations of $\delta g_{\mu\nu}$ under coordinate changes are defined by the invariance of the line element $ds^2$.  In particular, there is a coordinate freedom, called \emph{gauge freedom}, in mapping from the unperturbed spacetime to the perturbed spacetime.  For example, a given perturbed universe can be described using coordinates in which the spatial hypersurfaces of constant time coordinate have uniform total energy density, or alternatively by coordinates in which those hypersurfaces' normal curves have uniform local rate of expansion.  The two choices result in different matrices $\delta g_{\mu\nu}$.  Critical to the decomposition of Eq.~(\ref{delta_g}) is that the three classes of perturbations (scalar, vector, and tensor) do not mix with each other under the \emph{gauge transformation} from one choice of coordinates to another.

Concretely, a gauge transformation in the perturbed spacetime can be written as a change of coordinates
\begin{displaymath}
x^\mu\mapsto\hat{x}^\mu=x^\mu+\xi^\mu,
\end{displaymath}
which has four degrees of freedom.  So four out of the ten degrees of freedom of $\delta g_{\mu\nu}$ are unphysical gauge degrees.  The remaining six physical degrees of freedom may be decomposed into
\begin{itemize}
\item two scalar degrees, representing density perturbations;
\item two vector degrees, which are not of cosmological interest because they decay very quickly;
\item and two tensor degrees, representing gravitational radiation.
\end{itemize}
Of interest to us are the scalar degrees, which may be expressed as linear combinations of density (``adiabatic'') perturbations and entropy (``isocurvature'') perturbations~\cite{Bardeen}.  When the anisotropic stress vanishes (as is the case during matter domination, for example) the scalar sector reduces to a single physical degree of freedom.

Linear cosmological perturbation theory applies through most of the history of the Universe, and it applies today on sufficiently large scales; it breaks down only when and where perturbations grow in magnitude to order unity.

\subsection{Power spectra and cosmic variance}
The principal connection between cosmological perturbation theory and our observations of the perturbed universe is through the statistical power spectra of observable quantities.

Assuming statistical homogeneity and isotropy, a scalar perturbation $f$ on a given spatial hypersurface may be written as an isotropic Gaussian random scalar field on $\bm{R}^3$ with a power spectrum $P(k)$.  That is, if the Fourier transform of $f$ is given by
\begin{displaymath}
f(\bm{x})=\frac{1}{(2\pi)^{3/2}}\int d^3k\ \tilde{f}(\bm{k})e^{i\bm{k}\cdot\bm{x}},
\end{displaymath}
then for each $\bm{k}$, $\tilde{f}(\bm{k})$ is a Gaussian random variable with mean (over the statistical ensemble)
\begin{displaymath}
\left\langle\tilde{f}(\bm{k})\right\rangle=0
\end{displaymath}
and variance specified by
\begin{equation}
\label{statreln}
\left\langle\tilde{f}(\bm{k})\tilde{f}^\ast(\bm{k'})\right\rangle=\delta^3(\bm{k} - \bm{k'})P(k),
\end{equation}
depending only on $k\equiv|\bm{k}|$.  Note that this definition of $P(k)$ depends on the choice of normalization of the Fourier transform; the choice we have made is common in the literature (see Mukhanov~\cite[section 8.1]{Mukhanov} and Liddle and Lyth~\cite[section 4.3.2]{Liddle_Lyth}) and matches that used by the numerical tool \textsc{CAMB}~\cite{CAMB_notes}.

$P(k)$ has dimension $k^3\tilde{f}^2$, or $x^3f^2$.  In particular, if $f$ is dimensionless, then $P(k)$ has dimensions of volume, and it is therefore convenient to define a dimensionless power spectrum~\cite[\textit{ibidem}]{Liddle_Lyth}
\begin{displaymath}
\mathcal P(k)\equiv\frac{k^3}{2\pi^2}P(k).
\end{displaymath}

Critically, however, the power spectrum of $f$ depends on the spatial hypersurface we have chosen to define $f$.  If the hypersurface corresponds to a fixed coordinate time, then this choice depends on the gauge we are using.  For example, if $f$ is the matter density perturbation $\delta\rho_\text{m}/\rho_\text{m}$, then in the uniform matter density gauge, $\mathcal{P}(k)=0$ for all $k$.

Suppose we are interested in the variance of an estimate of a dimensionless cosmic parameter $F$ due to observation over a finite region $V$ of some spacelike hypersurface.  To estimate the mean value $\overline{F}$, we compute the mean $\widehat{F}_V$ over the volume $V$ of space in which we can observe $F$.  We want the statistical variance of this estimate.

Let us define a perturbation variable $f\equiv F-\overline{F}$.  Then the variance of the estimate $\widehat{F}_r$ (the mean over a spherical region of comoving radius $r$) is equal to the variance of $\widehat{f}_r$.  If the dimensionless power spectrum of $f$ is $\mathcal{P}(k)$, then (as we show in Appendix \ref{Scalar field variance}), this variance is simply
\begin{equation}
\label{Variance_equation}
\Var\left(\widehat{F}_r\right)=\left\langle\left|\widehat{f}_r\right|^2\right\rangle=\int\frac{dk}{k}\mathcal{P}(k)W_r^2(k),
\end{equation}
where we have defined the window function
\begin{displaymath}
W_r(k)\equiv3\frac{j_1(kr)}{kr}.
\end{displaymath}

\subsection{Gauge and cosmic parameters}
When we describe our Universe as a perturbed FLRW universe, the background FLRW spacetime has well-defined cosmological parameters which are spatially constant, including background values of $\Omega_X(\bm{x},t)$ (where $X$ here may stand for matter, radiation, curvature, or even dark energy), which we call $\overline{\Omega}_X$.  But gauge freedom tells us that there is no unique choice for the point identification map between the background FLRW universe and our perturbed universe; in other words, there is no unique description of the real Universe as a particular perturbed FLRW universe.  We have no preferred global choice of time coordinate, and since the local parameters $\Omega_X$ evolve over time, we see that perturbations from the background values $\overline{\Omega}_X$, which we will require to determine variances, are also gauge-dependent quantities.

It would seem as though gauge ambiguity leaves us with no place to plant our feet.  But the real problem is that in a perturbed FLRW universe, these background parameters are simply not well-defined quantities.

To escape gauge ambiguity, we need cosmic parameters which are independent of the time of observation for each FLRW universe.  To retain the utility of $\Omega_X$, the simplest choice is to pick the value of $\Omega_X$ at some physically fixed time.  The time we choose is that of last scattering, which at background level we may define as an instantaneous event taking place at redshift $z=1100$, and at perturbative level as a spacelike hypersurface of constant radiation energy density.\footnote{Last scattering actually occurs over an interval of time as the baryons gradually recombine.  We can define the time of last scattering more precisely to correspond to the maximum of the Thomson scattering visibility function.}  We denote the background value of $\Omega_X$ at last scattering by $\Omega_X^{\text{LS}}$.  As with conventional parameters $\Omega_X$, this definition carries over locally to the real perturbed Universe, although only on the hypersurface of last scattering itself.  We may extend the definition in a continuous fashion to points not on the last scattering hypersurface, for example by defining $\Omega_X^{\text{LS}}$ at any event to be the mean of the local $\Omega_X$ over the intersection of the hypersurface of last scattering with the light cone of that event, but this is not strictly necessary.

There are other reasons that it is a good choice to work on the last scattering surface.  At $z=1100$, nonlinear structure has not yet formed, and linear cosmological perturbation theory is valid on virtually all distance scales, particularly those which dominate the observable perturbation power spectra.  In addition, the contribution from dark energy is negligible at this time, so we need consider only matter, radiation, and curvature in our calculations.

$\Omega_X^{\text{LS}}$ is a good choice of parameter for one more reason.  We are interested in the theoretical limit of our ability to estimate cosmological parameters through electromagnetic observations.  Thus, consider an observer at event $O$ who can collect complete information from their causal past as far back as last scattering.  Let $\Sigma_O$ be the intersection of the causal past of $O$ with the hypersurface of last scattering; we call this the causal last scattering hypersurface of $O$ and denote it by $\Sigma_O$.  The region of spacetime observable by $O$ is the future Cauchy development of $\Sigma_O$, in which the physics is completely determined by information on $\Sigma_O$.  Thus, for a given component $X$, the observer's best estimate $\widehat{\Omega}_X^\text{LS}$ of the background value of $\Omega_X^\text{LS}$ is simply the mean of $\Omega_X$ over $\Sigma_O$,
\begin{equation}
\label{LSS_mean}
\widehat{\Omega}_X^\text{LS}\equiv\frac{1}{V(\Sigma_O)}\int_{\Sigma_O} \Omega_X\,dV.
\end{equation}
Note that, while the directly visible last scattering surface corresponds to just the {\em boundary} of $\Sigma_O$, we are interested here in {\em ultimate} limits.  The interior of $\Sigma_O$ can be observed, for example, by mapping the distribution of matter with galaxy or $21\,\text{cm}$ line observations.

We can then compute the cosmic variance of the estimate $\widehat{\Omega}_X^\text{LS}$ using Eq.~(\ref{Variance_equation}), if we know the power spectrum of the perturbations $\delta\Omega_X$ of $\Omega_X$ on the hypersurface of last scattering.

\section{\label{PLS}Perturbations at last scattering}
The evolution of a realistic set of coupled cosmological perturbations, the crux of cosmological perturbation theory, is complicated.  Fortunately there exist several numerical tools which allow us to obtain power spectra.  In this section we use \textsc{CAMB}~\cite{lcl00},\footnote{\textsc{CAMB} is available at \href{http://camb.info/}{http://camb.info/}.} which performs calculations in the gauge in which the normals to the constant-time slices are comoving with cold dark matter (CDM).  Since CDM is assumed not to interact with the other matter components, this gauge is also a synchronous gauge.

The constant-time hypersurfaces of the CDM-comoving gauge are not hypersurfaces of constant radiation energy density, which defines the hypersurface of last scattering.  At linear order, the transformation of a four-scalar between these two slices is given by the first term in a Taylor expansion in proper time $t$.  Writing $X$ for scalar quantities on the CDM-comoving hypersurface and $\widetilde{X}$ for those quantities on the hypersurface of last scattering (note that we are now changing the meaning of the tilde), we have
\begin{equation}
\label{Taylor_expansion}
\widetilde{X}(\bm{x})=X(\bm{x})+\Delta t(\bm{x})\partial_t X(\bm{x}),
\end{equation}
where $\Delta t$ is the proper time displacement between the two slices.  This transformation applies to an energy density, $X = \rho$, even though $\rho$ is not a true four-scalar, since at linear order the density does not change under boosts.  Similarly, it applies to the expansion rate, $X = H$, when we keep fixed the worldlines whose expansion $H$ measures.

The last scattering hypersurface is a hypersurface of constant radiation energy density.  Assuming adiabatic perturbations, this is also a hypersurface of constant matter density on large (super-Hubble) scales.  We will see later that most of the contribution to the variances we calculate comes from scales that are super-Hubble at last scattering, so that it is a reasonable approximation to take last scattering to occur at constant matter density.  Letting $X=\rho_\text{m}$, the left hand side of Eq.~(\ref{Taylor_expansion}) is constant in $\bm{x}$.  For the right hand side also to be constant, we require
\begin{displaymath}
\Delta t=-\frac{\delta\rho_\text{m}(\bm{x})}{\partial_t\rho_\text{m}(\bm{x})},
\end{displaymath}
where $\delta\rho_\text{m}$ is in CDM-comoving gauge.  Taking expectations, Eq.~(\ref{Taylor_expansion}) now gives
\begin{displaymath}
\overline{\widetilde{X}}-\overline X=\left\langle\Delta t\,\partial_t X\right\rangle=-\left\langle\delta\rho_\text{m}\right\rangle\frac{\partial_t\overline X}{\partial_t\overline{\rho}_\text{m}}=0,
\end{displaymath}
where we have replaced $\partial_t X$ and $\partial_t\rho_\text{m}$ by the mean values $\partial_t\overline{X}$ and $\partial_t\overline{\rho}_\text{m}$, since we are working to first order in $\delta\rho_\text{m}$.  We may therefore convert Eq.~(\ref{Taylor_expansion}) into a form which applies to perturbations $\delta X\equiv X-\overline{X}$:
\begin{displaymath}
\delta\widetilde{X}(\bm{x})=\delta X(\bm{x})-\frac{\delta\rho_\text{m}(\bm{x})}{\partial_t\rho_\text{m}(\bm{x})}\partial_t\overline X.
\end{displaymath}
Again replacing $\partial_t\rho_\text{m}$ by $\partial_t\overline{\rho}_\text{m}$, we take the Fourier transform to obtain
\begin{equation}
\label{Taylor_expansion_delta}
\delta\widetilde{X}(\bm{k})=\delta X(\bm{k})-\frac{\delta\rho_\text{m}(\bm{k})}{\partial_t\overline{\rho}_\text{m}}\partial_t\overline X.
\end{equation}

In what follows, we will work exclusively in Fourier space.

\subsection{Matter density parameter perturbations\texorpdfstring{ $\delta\Omega_\text{m}$}{}}
From Eq.~(\ref{omegamdef}), it is easy to work out a relationship between $\delta\Omega_\text{m}$, $\delta\rho_\text{m}$, and $\delta H$:
\begin{equation}
\label{delta_Omega_m}
\frac{\delta\Omega_\text{m}}{\overline{\Omega}_\text{m}}=\frac{\delta\rho_\text{m}}{\overline{\rho}_\text{m}}-2\frac{\delta H}{\overline{H}}.
\end{equation}

To calculate the perturbations to $\Omega_\text{m}$ on the last scattering hypersurface, we first evaluate this equation on a CDM-comoving hypersurface and then use Eq.~(\ref{Taylor_expansion_delta}) to switch to a surface of constant energy density.  This is valid since, as explained above, both $\rho_\text{m}$ and $H$ can be treated as four-scalars.

The first thing to do is to express $\delta H$ in terms of $\delta\rho_\text{m}$.  Using the continuity equation
\begin{displaymath}
\partial_t\rho_\text{m} = -3H\rho_\text{m},
\end{displaymath}
which holds to linear order when $H$ measures the expansion rate of the comoving matter worldlines, we find
\begin{align*}
\delta H&=H-\overline H\\*
&=-\frac{1}{3}\left(\frac{\partial_t\overline{\rho}_\text{m}+\partial_t\delta\rho_\text{m}}{\overline{\rho}_\text{m}+\delta\rho_\text{m}}-\frac{\partial_t\overline{\rho}_\text{m}}{\overline{\rho}_\text{m}}\right)\\*
&=-\frac{1}{3}\partial_t\left(\frac{\delta\rho_\text{m}}{\overline{\rho}_\text{m}}\right),
\end{align*}
dropping terms in $\left(\delta\rho_\text{m}\right)^2$ along the way.

The time derivative here cannot be found analytically in general, since the perturbation equations are second-order in $\rho_\text{m}$.  However, as we will show in Appendix~\ref{CAMB_details}, this information can be extracted numerically from \textsc{CAMB}.

Finally, we employ Eq.~(\ref{Taylor_expansion_delta}) to move to the last scattering hypersurface:
\begin{displaymath}
\frac{\delta\widetilde{\Omega}_\text{m}}{\overline{\Omega}_\text{m}}=\frac{\delta\Omega_\text{m}}{\overline{\Omega}_\text{m}}-\frac{\delta\rho_\text{m}}{\partial_t\overline{\rho}_\text{m}}\frac{\partial_t\overline{\Omega}_\text{m}}{\overline{\Omega}_\text{m}}.
\end{displaymath}
Using the background relation $\overline{\Omega}_\text{m}+\overline{\Omega}_\text{r}=1$ and the continuity equations for $\rho_\text{m}$ and $\rho_\text{r}$, we can rewrite the background component of the second term as
\begin{displaymath}
\frac{1}{\partial_t\overline{\rho}_\text{m}}\frac{\partial_t\overline{\Omega}_\text{m}}{\overline{\Omega}_\text{m}}=-\frac{1}{3}\frac{\overline{\Omega}_\text{r}}{\overline{\rho}_\text{m}}.
\end{displaymath}

Putting this all together, we have
\begin{equation}
\label{my_awesome_equation}
\frac{\delta\widetilde{\Omega}_\text{m}}{\overline{\Omega}_\text{m}}=\left(1+\frac{1}{3}\overline{\Omega}_\text{r}\right)\frac{\delta\rho_\text{m}}{\overline{\rho}_\text{m}}+\frac{2}{3\overline{H}}\partial_t\left(\frac{\delta\rho_\text{m}}{\overline{\rho}_\text{m}}\right).
\end{equation}
This equation relates the amplitude of perturbations in $\Omega_\text{m}$ on the hypersurface of last scattering to the amplitude of perturbations in $\rho_\text{m}$ on the nearby CDM-comoving hypersurface used by \textsc{CAMB}.

If we make the simplifying assumption of matter domination at last scattering, $\overline{\Omega}_\text{m}=1$, then we have the well-known result $\delta\rho_\text{m}/\rho_\text{m}\propto a$ (see, for example, Press and Vishniac~\cite[Eqs.~32c and~9]{Press_Vishniac}), so
\begin{displaymath}
\partial_t\left(\frac{\delta\rho_\text{m}}{\overline{\rho}_\text{m}}\right)=\overline{H}\frac{\delta\rho_\text{m}}{\overline{\rho}_\text{m}}.
\end{displaymath}
Also, the correction term for moving to the hypersurface of constant density vanishes.  In this case, Eq.~(\ref{my_awesome_equation}) is simply
\begin{displaymath}
\frac{\delta\widetilde{\Omega}_\text{m}}{\overline{\Omega}_\text{m}}=\frac{5}{3}\frac{\delta\rho_\text{m}}{\overline{\rho}_\text{m}}.
\end{displaymath}

\subsection{Radiation density parameter perturbations\texorpdfstring{ $\delta\Omega_\text{r}$}{}}
We can also examine the variance of $\Omega_\text{r}$, since Eq.~(\ref{delta_Omega_m}) still holds if we substitute radiation for matter:
\begin{displaymath}
\frac{\delta\Omega_\text{r}}{\overline{\Omega}_\text{r}}=\frac{\delta\rho_\text{r}}{\overline{\rho}_\text{r}}-2\frac{\delta H}{\overline{H}}.
\end{displaymath}
On the hypersurface of last scattering, adiabaticity tells us that $\delta\widetilde{\rho}_\text{r}=\delta\widetilde{\rho}_\text{m}=0$ (on large scales), so
\begin{equation}
\label{Omega_r_Omega_m}
\frac{\delta\widetilde{\Omega}_\text{r}}{\overline{\Omega}_\text{r}}=-2\frac{\delta\widetilde{H}}{\overline{H}}=\frac{\delta\widetilde{\Omega}_\text{m}}{\overline{\Omega}_\text{m}}.
\end{equation}
(Recall that in defining both $\Omega_\text{m}$ and $\Omega_\text{r}$, we take $H$ to be the expansion rate of the same comoving matter worldlines.)

\subsection{The curvature parameter\texorpdfstring{ $\Omega_K$}{}}
\label{tpwcurvsec}
The spatial curvature scalar ${}^{(3)}\!R$, for hypersurfaces orthogonal to the matter comoving worldlines, enters the local energy constraint (Friedmann) equation as (see, e.g., Ref.~\cite[Eqs~4.165 and~14.132]{Liddle_Lyth})
\begin{displaymath}
H^2=\frac{8\pi G}{3}\left(\rho_\text{m}+\rho_\text{r}\right)-\frac{{}^{(3)}\!R}{6}.
\end{displaymath}
Thus, even with our local definitions of the parameters, we have the familiar relation $\Omega_\text{m}+\Omega_\text{r}+\Omega_K=1$.  Also, Eq.~(\ref{Taylor_expansion_delta}) assures us that with $\overline{\Omega}_K=0$, $\Omega_K$ is independent of the slicing we are working on.  However, it is simplest to continue working on the constant energy density hypersurface, obtaining
\begin{equation}
\label{Omega_K}
\widetilde{\Omega}_K=1-\widetilde{\Omega}_\text{m}-\widetilde{\Omega}_\text{r}=-\frac{\delta\widetilde{\Omega}_\text{m}}{\overline{\Omega}_\text{m}}
\end{equation}
using Eq.~(\ref{Omega_r_Omega_m}).

It is important to note that, regardless of which slicing we are using, $\Omega_K$ is defined in terms of \emph{comoving curvature}, that is, the spatial curvature of hypersurfaces orthogonal to the matter-comoving worldlines.

\section{\label{numerical_section}The variances of \texorpdfstring{$\widehat{\Omega}_\text{m}^\text{LS}$, $\widehat{\Omega}_\text{r}^\text{LS}$, and $\widehat{\Omega}_K^\text{LS}$}{the parameters}}
In this section, we drop the use of the bar to denote mean values, in order to avoid cumbersome notation.

To calculate numerical values for the variances, we must choose a set of background cosmological parameters.  We use the following maximum-likelihood values from the five-year WMAP data combined with baryon acoustic oscillations and Type Ia supernovae~\cite[Table 1]{WMAP5}:
\begin{align*}
\Omega_{\text{b},0}h_0^2&=0.02263\\*
\Omega_{\text{m},0}h_0^2&=0.1362\\*
h_0&=0.703\\*
\mathcal{P}_\mathcal{R}(k_0)&=2.42\times10^{-9}\\*
n_s&=0.961
\end{align*}
Here subscript $0$ refers to values today, $h$ is the Hubble rate in units of $100\,\text{km}\,\text{s}^{-1}\,\text{Mpc}^{-1}$, $\mathcal{P}_\mathcal{R}(k)$ is the power spectrum of the primordial comoving curvature perturbation $\mathcal{R}$, and WMAP uses a pivot scale $k_0 = 0.002\,\text{Mpc}^{-1}$.  The scalar index $n_s$ determines the departure from scale invariance in the primordial spectrum via
\begin{displaymath}
\mathcal{P}_\mathcal{R}(k)=\mathcal{P}_\mathcal{R}(k_0)\left(\frac{k}{k_0}\right)^{n_s - 1}.
\end{displaymath}
The corresponding value of $\Omega_{\text{m},0}$ is $0.276$.  Since current data are consistent with (and the inflationary model predicts) a flat $\Omega_K=0$ universe, we use this value for simplicity.  We also have $\Omega_{\text{r},0}\approx4.17\times10^{-5}h_0^{-2}$~\cite[section 2.2.2]{Liddle_Lyth}.  Together, these give $\Omega_{\Lambda,0}=0.724$.

With these parameter values, the comoving distance to last scattering is $r_\text{LS}\approx9.86h^{-1}\,\text{Gpc}\approx14.0\,\text{Gpc}$, and at last scattering we have $\Omega_\text{r}^\text{LS}\approx0.252$ and $\Omega_\text{m}^\text{LS}\approx0.748$, so it is not adequate to assume matter domination at last scattering (see Appendix \ref{last_scattering} for the details).

Using Eqs.~(\ref{Variance_equation}) and~(\ref{my_awesome_equation}) together with data produced using \textsc{CAMB} with the above parameter values as input (see Appendix~\ref{CAMB_details} for the details), we have computed the fractional variance of $\widehat{\Omega}_\text{m}^\text{LS}$ to be
\begin{displaymath}
\Var\left(\frac{\widehat{\delta\Omega}\vphantom{\Omega}_\text{m}^\text{LS}}{\Omega_\text{m}^\text{LS}}\right)=1.51\times10^{-15}.
\end{displaymath}
The absolute variance of $\widehat{\Omega}_\text{m}^\text{LS}$ is therefore
\begin{displaymath}
\Var\left(\widehat{\Omega}_\text{m}^\text{LS}\right)=\left(\Omega_\text{m}^\text{LS}\right)^2\Var\left(\frac{\widehat{\delta\Omega}\vphantom{\Omega}_\text{m}^\text{LS}}{\Omega_\text{m}^\text{LS}}\right)=8.44\times10^{-16}.
\end{displaymath}
Its standard deviation, which we shall call $\sigma_\text{m}^\text{LS}$, is
\begin{displaymath}
\sigma_\text{m}^\text{LS}\equiv\sqrt{\Var\left(\widehat{\Omega}_\text{m}^\text{LS}\right)}=2.90\times10^{-8}.
\end{displaymath}

By Eq.~(\ref{Omega_r_Omega_m}), the fractional variance of $\widehat{\Omega}_\text{r}^\text{LS}$ is equal to that of $\widehat{\Omega}_\text{m}^\text{LS}$; its standard deviation is thus
\begin{displaymath}
\sigma_\text{r}^\text{LS}=\Omega_\text{r}^\text{LS}\sqrt{\Var\left(\frac{\widehat{\delta\Omega}\vphantom{\Omega}_\text{m}^\text{LS}}{\Omega_\text{m}^\text{LS}}\right)}=9.79\times10^{-9}.
\end{displaymath}

Finally, using Eq.~(\ref{Omega_K}), we see that the absolute variance of $\widehat{\Omega}_K^\text{LS}$ is equal to the fractional variance of $\widehat{\Omega}_\text{m}^\text{LS}$, so its standard deviation is
\begin{equation}
\label{tpwsigmak}
\sigma_K^\text{LS}=3.88\times10^{-8}.
\end{equation}

Note that $\sigma_\text{m}^\text{LS}+\sigma_\text{r}^\text{LS}=\sigma_K^\text{LS}$, since perturbations in matter and radiation are perfectly correlated by the assumption of adiabaticity.

\section{\label{directOmegaKsec}Direct calculation for \texorpdfstring{$\Omega_K$}{Omega\_K}}
In the preceding sections we arrived at a result for the variance of $\widehat{\Omega}_K^\text{LS}$ via a calculation for $\Omega_\text{m}$ and the Einstein energy constraint equation.  It is possible to take an alternative and more direct route to $\sigma_K^\text{LS}$, working straight from the definition Eq.~(\ref{omegakdef}).  The idea is simply to calculate the spatial curvature of the last scattering hypersurface itself, i.e.\ the curvature of a hypersurface of uniform radiation density, then average over the (in principle) observable volume $\Sigma_O$ using Eq.~(\ref{LSS_mean}), and finally take the variance.

The spatial Ricci curvature ${}^{(3)}\!R$ of the last scattering hypersurface is related to the metric curvature perturbation $\psi_\gamma$ (the isotropic scalar part of the perturbed spatial metric) by~\cite[Eq.~14.129]{Liddle_Lyth}
\begin{equation}
{}^{(3)}\!R = -4\frac{k^2}{a^2}\psi_\gamma,
\end{equation}
where $\psi_\gamma$ is the scalar $\psi$ of Eq.~(\ref{delta_g}), evaluated in the uniform radiation density gauge (here radiation refers only to photons, not neutrinos, hence the subscript $\gamma$).  Since we wish to write the variance in terms of the power spectrum of the primordial comoving curvature perturbation $\mathcal{R}$, we define a transfer function $T_K(k)$ through
\begin{equation}
\psi_\gamma(k,t_\text{LS}) = T_K(k) \mathcal{R}(k).
\end{equation}
Combining these expressions with Eq.~(\ref{omegakdef}) gives
\begin{equation}
\Omega_{K,\gamma}(t_\text{LS})=\frac{2}{3}\left(\frac{k}{a_\text{LS}H_\text{LS}}\right)^2T_K(k) \mathcal{R}(k),
\end{equation}
for the curvature parameter at the time of last scattering.  Finally, averaging over the sphere $\Sigma_O$ and taking the variance using Eq.~(\ref{Variance_equation}), we find
\begin{equation}
\label{jzvarkint}
\Var\left(\widehat{\Omega}_{K,\gamma}^\text{LS}\right) = \frac{4}{9}\int\frac{dk}{k}\left(\frac{k}{a_\text{LS}H_\text{LS}}\right)^4T_K^2(k)\mathcal{P}_\mathcal{R}(k)W_{r_\text{LS}}^2(k).
\end{equation}

Now all that is required is to determine the transfer function $T_K(k)$ for the uniform radiation gauge curvature perturbation.  Such detailed information can be readily extracted from the software package \textsc{COSMICS}~\cite{mabe}.\footnote{\textsc{COSMICS} is available at \href{http://web.mit.edu/edbert/}{http://web.mit.edu/edbert/}.}  Extracting the transfer function and performing the integral in Eq.~(\ref{jzvarkint}), we find standard deviation
\begin{equation}
\label{jzsigmak}
\sigma_{K,\gamma}^\text{LS} = 4.35\times10^{-8}.
\end{equation}

To understand the apparent disparity between this result and that of Eq.~(\ref{tpwsigmak}) above, recall from section~\ref{tpwcurvsec} that the curvature parameter calculated there describes the spatial curvature of hypersurfaces orthogonal to the comoving worldlines.  These comoving slices coincide with uniform total energy slices on super-Hubble scales~\cite{wmll00}, and hence with uniform radiation slices for adiabatic modes.  The integrand in Eq.~(\ref{jzvarkint}) receives most of its support from scales between the Hubble scale at last scattering and scales about $100$ times larger, so we expect the two estimates to be close.  However, there is a significant contribution from Hubble-scale modes, which largely accounts for the discrepancy.  Using \textsc{COSMICS}, we can repeat the above calculation, but computing the curvature of the comoving slice at last scattering instead of the uniform radiation density slice.  The result is
\begin{equation}
\label{jzsigmakr}
\sigma_{K,\mathcal{R}}^\text{LS} = 3.74\times10^{-8},
\end{equation}
which is now very close to Eq.~(\ref{tpwsigmak}).  We believe the remaining disparity is due to the fact that the CDM-comoving gauge used by \textsc{CAMB} does not coincide precisely with the total matter comoving gauge.

\section{\label{IaD}Interpretation and discussion}
Our calculations in the preceding sections are all evaluated at the time of last scattering.  Thus they are not directly comparable with standard determinations of the density parameters $\Omega_{\text{m},0}$, $\Omega_{\text{r},0}$, and $\Omega_{K,0}$, which are evaluated today.  For a homogeneous background spatial curvature, it is simple to relate the curvature parameter $\Omega_K$ at different times using Eq.~(\ref{omegaktdep}).  Namely, $\Omega_K$ is proportional to the square of the comoving Hubble length, since the comoving spatial curvature length remains constant.  For the inhomogeneous case, although variances are perfectly well-defined on the hypersurface of last scattering, we are again confronted with ambiguity in how to translate them to values today.  We choose to translate the standard deviation of $\widehat{\Omega}_K$, Eq.~(\ref{tpwsigmak}), in the same way as a background curvature, using Eq.~(\ref{omegaktdep}) to obtain
\begin{equation}
\label{sigmaktoday}
\sigma_K^0\equiv\left(\frac{a_\text{LS}H_\text{LS}}{a_0H_0}\right)^2\sigma_K^\text{LS} = 1.58\times10^{-5},
\end{equation}
where we have used the Friedmann equation to determine $H_\text{LS}$.

We can similarly translate the results from section~\ref{directOmegaKsec} to today.  For the spatial curvature of the last scattering hypersurface itself, i.e.\ the uniform radiation density slice, Eq.~(\ref{jzsigmak}), we find
\begin{equation}
\sigma_{K,\gamma}^0 = 1.77\times10^{-5}.
\end{equation}
Translating the curvature of comoving slices, Eq.~(\ref{jzsigmakr}), to today gives
\begin{equation}
\label{jzsigkr0}
\sigma_{K,\mathcal{R}}^0 = 1.52\times10^{-5},
\end{equation}
which again is very close to the estimate Eq.~(\ref{sigmaktoday}).

\begin{figure}[t]
\includegraphics[width=\columnwidth]{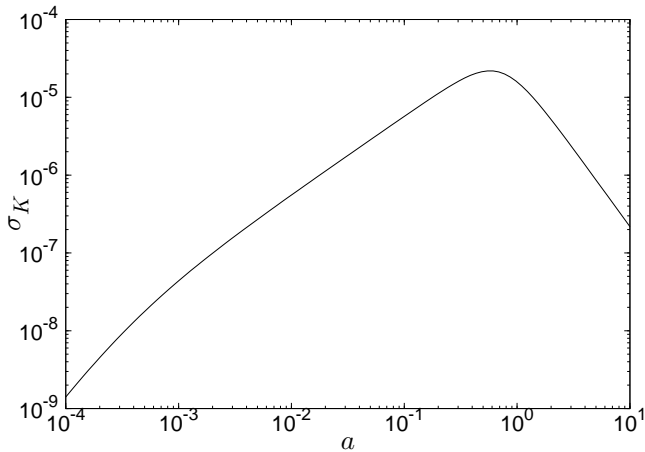}
\caption{\label{sigma_K_time}The time dependence of $\sigma_K$.  The change of slope is due to the transitions from radiation domination to matter domination and from matter domination to dark energy domination.  The standard deviation today has already begun to decrease as we approach de~Sitter.}
\end{figure}

To help understand this evolution, it may be helpful to consider what happens to the variance of $\widehat{\Omega}_K$ at \emph{future} times, again using the prescription of Eq.~(\ref{omegaktdep}).  The variance will approach zero as we approach the late time de~Sitter stage,\footnote{Here we are assuming for simplicity that the dark energy is a pure cosmological constant.} when the comoving Hubble length decreases indefinitely; see Fig.~\ref{sigma_K_time}.  This is a reflection of precisely the same mechanism by which inflation drives spatial curvature towards zero in the early universe.  Indeed, the variance calculated for today in Eqs.~(\ref{sigmaktoday}) to~(\ref{jzsigkr0}) has already begun to decrease significantly, as we are already approaching $\Lambda$-domination.  However, it could be argued that this decrease in the variance of $\widehat{\Omega}_K$ at late times is deceptive, in that we are simply scaling essentially the same last-scattering variance to a smaller and smaller comoving Hubble length.  ($r_\text{LS}$ will increase somewhat in the future, but it will not grow without bound.)  Recall that the large-scale comoving curvature perturbations are described by a \emph{constant} amplitude of order $10^{-5}$.  An alternative prescription to Eq.~(\ref{omegaktdep}) for evolving the curvature variance would be to scale it with $r_\text{LS}$ rather than the Hubble length.  Then the variance in $\widehat{\Omega}_K$ would approach a constant at late times, of order $10^{-5}$.

In summary, we have quantified the common lore that, within the standard inflationary paradigm, we can never measure $\Omega_K$ to better precision than of order $10^{-5}$.  Our calculations are completely independent of the details of any particular observational technique and hence provide a fundamental limit for any measurement of curvature.  In particular, it will be impossible to determine the sign of $\Omega_K$, and thus whether the Universe appears to be open or closed, if $\left|\Omega_{K,0}\right|\lesssim\sigma_K^0\sim10^{-5}$.  On the other hand, if we some day determine $\left|\Omega_{K,0}\right|\gg\sigma_K^0$, then we will know that (within the framework of the Standard Model) we have measured the very large scale geometry of our Universe and not merely a local fluctuation.  While current observational constraints on curvature are well above this fundamental limit, the mapping of a significant portion of our past light cone, as would take place with planned $21\,\text{cm}$ hydrogen line experiments~\cite{0802.1710}, will bring us close to our ultimate ability to constrain curvature, matter content, and other cosmological parameters.

\begin{acknowledgments}
The authors thank Douglas Scott for several helpful discussions and Rainer Dick for proofreading and useful feedback.  This research was supported in part by the Natural Sciences and Engineering Research Council of Canada and in part by grant RFP1-06-006 from the Foundational Questions Institute.  Research at Perimeter Institute is supported by the Government of Canada through Industry Canada and by the Province of Ontario through the Ministry of Research and Innovation.
\end{acknowledgments}

\appendix

\section{\label{Scalar field variance}Scalar field variance}
Consider the mean value of a function $f\colon\bm{R}^3\rightarrow\bm{R}$ over some spherical volume $\frac{4}{3}\pi r^3$ (``top-hat'') of flat space.  We will denote this quantity by $\widehat{f}_r$:
\begin{align*}
\widehat{f}_r&=\frac{1}{\frac{4}{3}\pi r^3}\int_{|\bm{x}|<r}d^3x\ f(\bm{x})\\*
&=\frac{1}{\frac{4}{3}\pi r^3}\int_{|\bm{x}|<r}d^3x\frac{1}{(2\pi)^{3/2}}\int d^3k\ \tilde{f}(\bm{k})e^{i\bm{k}\cdot\bm{x}}\\*
&=\frac{3}{(2\pi)^{3/2}}\int d^3k\ \tilde{f}(\bm{k})\frac{j_1(kr)}{kr}.
\end{align*}
Here $j_1$ is the spherical Bessel function of first order,
\begin{displaymath}
j_1(x)=\frac{\sin x-x\cos x}{x^2}.
\end{displaymath}

The variance of $\widehat{f}_r$ is
\begin{displaymath}
\left\langle\left|\widehat{f}_r\right|^2\right\rangle-\left|\left\langle\widehat{f}_r\right\rangle\right|^2.
\end{displaymath}
If $f$ is a perturbation, then $\left\langle\widehat{f}_r\right\rangle=\left\langle\widehat{f}\right\rangle=0$, and thus the variance of $\widehat{f}_r$ is simply
\begin{align*}
\left\langle\left|\widehat{f}_r\right|^2\right\rangle&=\left\langle\left|\frac{3}{(2\pi)^{3/2}}\int d^3k\ \tilde{f}(\bm{k})\frac{j_1(kr)}{kr}\right|^2\right\rangle\\*
&=\frac{1}{(2\pi)^3}\int d^3k\ \left\langle\left|\tilde{f}(\bm{k})\right|^2\right\rangle\left(3\frac{j_1(kr)}{kr}\right)^2\\*
&=\frac{1}{(2\pi)^3}\int 4\pi k^2dk\ P(k)\left(3\frac{j_1(kr)}{kr}\right)^2\\*
&=\int\frac{dk}{k}\mathcal{P}(k)\left(3\frac{j_1(kr)}{kr}\right)^2,
\end{align*}
where we have used Eq.~(\ref{statreln}).

\section{\label{CAMB_details}\textsc{CAMB} details}

Here we provide some details of our numerical implementation for the benefit of readers who wish to reproduce the results for different models.

\textsc{CAMB} produces two types of data files.

The \texttt{matterpower} files are two-column files in which the first column gives values of $k/h_0$ and the second column gives corresponding values of $h_0^3P(k)$, where $h_0$ is the dimensionless Hubble parameter evaluated today and $P(k)$ is the dimensional power spectrum of the quantity $\delta\rho_\text{m}/\rho_\text{m}$ in the CDM-comoving gauge.

The \texttt{transfer} files have seven columns, in which the first column is again $k/h_0$ and the remaining columns contain internally-defined dimensional transfer functions.  We are interested in the total matter transfer function, which is found in the seventh column.  We will call it $T_\text{C}(k)$ to distinguish it from the dimensionless transfer function $T(k)$ commonly used in the literature.

$T_\text{C}$ is related to the matter power spectrum $P(k)$ by
\begin{displaymath}
P(k)=T_\text{C}^2(k)\cdot2\pi^2k\mathcal{P}_\mathcal{R}(k),
\end{displaymath}
where $\mathcal{P}_\mathcal{R}(k)$ is the dimensionless power spectrum of the primordial comoving curvature perturbation, which can be constrained observationally.  (For a scale-invariant primordial spectrum, $\mathcal{P}_\mathcal{R}(k) = \text{const}$.)  Thus we have the convenient relation
\begin{equation}
\label{CAMB_transfer}
\mathcal{P}(k)=k^4T_\text{C}^2(k)\mathcal{P}_\mathcal{R}(k).
\end{equation}
From this, we see that
\begin{displaymath}
\frac{\delta\rho_\text{m}}{\overline{\rho}_\text{m}}\propto k^2T_\text{C}(k),
\end{displaymath}
so
\begin{displaymath}
\partial_t\left(\frac{\delta\rho_\text{m}}{\overline{\rho}_\text{m}}\right)=\frac{\partial_t T_\text{C}(k)}{T_\text{C}(k)}\frac{\delta\rho_\text{m}}{\overline{\rho}_\text{m}}.
\end{displaymath}

We may therefore rewrite Eq.~(\ref{my_awesome_equation}) as
\begin{equation}
\label{CAMB_awesome_equation}
\frac{\delta\widetilde{\Omega}_\text{m}}{\overline{\Omega}_\text{m}}=\Gamma(k)\frac{\delta\rho_\text{m}}{\overline{\rho}_\text{m}},
\end{equation}
where we have defined
\begin{displaymath}
\Gamma(k)\equiv1+\frac{1}{3}\overline{\Omega}_\text{r}+\frac{2}{3\overline{H}}\frac{\partial_t T_\text{C}(k)}{T_\text{C}(k)}.
\end{displaymath}
Hence, we have modified the \textsc{CAMB} source code to output $\partial_\tau a$ and $\partial_\tau T_\text{C}(k)$ as well (where $\tau$ is conformal time), allowing us to compute $\Gamma(k)$ numerically.

Inserting Eqs.~(\ref{CAMB_transfer}) and~(\ref{CAMB_awesome_equation}) into Eq.~(\ref{Variance_equation}), we see that the fractional variance of $\widehat{\Omega}_\text{m}^\text{LS}$ is
\begin{displaymath}
\Var\left(\frac{\widehat{\delta\Omega}\vphantom{\Omega}_\text{m}^\text{LS}}{\Omega_\text{m}^\text{LS}}\right)=\int\frac{dk}{k}\Gamma^2(k)k^4T_\text{C}^2(k)\mathcal{P}_\mathcal{R}(k)W_{r_\text{LS}}^2(k).
\end{displaymath}
To evaluate this expression numerically, it is not sufficient simply to perform a Riemann sum, since the window function $W_{r_\text{LS}}(k)$ oscillates on a scale of $r_\text{LS}^{-1}\sim10^{-4}\,\text{Mpc}^{-1}$, but \textsc{CAMB} data points can be separated by up to $\sim10^{-2}\,\text{Mpc}^{-1}$.  (The function $T_\text{C}(k)$ varies much more slowly with $k$ than does $W_{r_\text{LS}}(k)$.)  The solution is to construct linear interpolations of $\Gamma(k)T_\text{C}(k)$ and then to perform a piecewise integral, including tail terms.  We have done this on a set of data produced by \textsc{CAMB} using the parameter values given in section~\ref{numerical_section} and with the \textsc{CAMB} parameters \texttt{transfer\_kmax} set to $1$ and \texttt{accuracy\_boost} set to $3$.

\begin{figure}[t]
\includegraphics[width=\columnwidth]{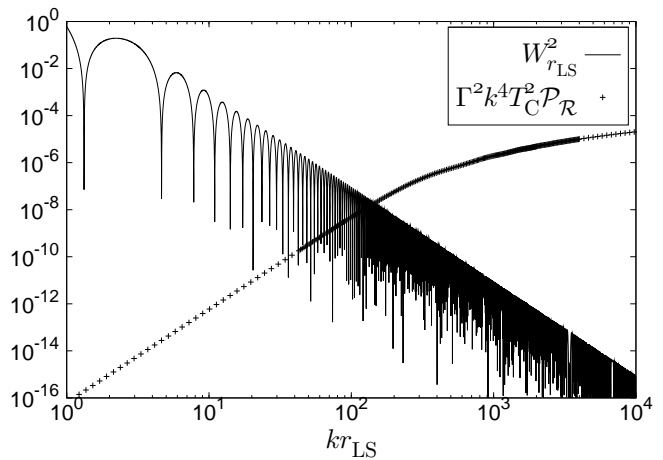}
\caption{\label{integrand_plot}The integrand components $W_{r_\text{LS}}^2$ and $\Gamma^2k^4T_\text{C}^2\mathcal{P}_\mathcal{R}$.  Note the rapid oscillations of $W$ relative to the spacing of the data points.}
\end{figure}

We observe that typical values of $\Gamma(k)$ are close to $2$, representing a correction of $20\%$ to the matter-domination value of $5/3$.  The integrand components $W_{r_\text{LS}}^2$ and $\Gamma^2k^4T_\text{C}^2\mathcal{P}_\mathcal{R}$ are plotted in Fig.~\ref{integrand_plot}.

\section{\label{last_scattering}Parameters for last scattering}
To estimate $\Omega_\text{r}$ at last scattering, we simply work backward from today:
\begin{align*}
\Omega_\text{r}^\text{LS}&=\frac{\rho_\text{r,0}a_\text{LS}^{-4}}{\rho_\text{r,0}a_\text{LS}^{-4}+\rho_\text{m,0}a_\text{LS}^{-3}+\rho_\Lambda}\\*
&=\left(1+\frac{\Omega_\text{m,0}}{\Omega_\text{r,0}}a_\text{LS}+\frac{\Omega_{\Lambda,0}}{\Omega_\text{r,0}}a_\text{LS}^4\right)^{-1}\\*
&\approx0.252,
\end{align*}
with $a_0 \equiv 1$ today.  Similarly, we find
\begin{displaymath}
\Omega_\text{m}^\text{LS}\approx0.748,
\end{displaymath}
and $\Omega_\Lambda$ at last scattering is negligible.

To calculate the comoving distance to last scattering, we start with the form of the Friedmann equation
\begin{displaymath}
\left(\frac{\dot{a}}{a}\right)^2=H_0^2\left(\Omega_{\text{r},0}a^{-4}+\Omega_{\text{m},0}a^{-3}+\Omega_{\Lambda,0}\right)
\end{displaymath}
in terms of density parameters evaluated today, with the dot denoting differentiation with respect to proper time.

Therefore the distance to last scattering is
\begin{align*}
r_\text{LS}&=\int_{t_\text{LS}}^{t_0}\frac{dt}{a(t)}=\int_{a_\text{LS}}^{a_0}\frac{da}{a\dot{a}}\\*
&=\frac{1}{H_0}\int_\frac{1}{1+z_\text{LS}}^1\frac{da}{\sqrt{\Omega_{\text{r},0}+\Omega_{\text{m},0}a+\left(1-\Omega_{\text{r},0}-\Omega_{\text{m},0}\right)a^4}}\\*
&\approx9.86h^{-1}\,\text{Gpc}.
\end{align*}

\bibliography{cosmic_variance_of_omega}

\end{document}